\documentclass[12pt,american]{article}
\usepackage[utf8]{inputenc}
\usepackage[a4paper]{geometry}
\usepackage{babel}
\usepackage{amsmath}
\usepackage{amssymb}
\usepackage{setspace}
\usepackage{esint}
\usepackage[authoryear]{natbib}
\PassOptionsToPackage{normalem}{ulem}
\usepackage{ulem}
\usepackage[unicode=true,pdfusetitle,
 bookmarks=true,bookmarksnumbered=false,bookmarksopen=false,
 breaklinks=false,pdfborder={0 0 1},backref=false,colorlinks=false]
 {hyperref}

\usepackage{bm}
\usepackage{enumitem}
\usepackage{multirow}

\makeatletter

\providecommand{\tabularnewline}{\\}

\@ifundefined{date}{}{\date{}}

\usepackage{indentfirst}
\usepackage{babel}

\usepackage{graphicx}
\usepackage{amsmath}


\begin{document}

\title{Empirical Likelihood Based Summary ROC Curve for Meta-Analysis of Diagnostic Studies}

\author{ShengLi Tzeng\\
 Institute of Statistical Science\\
Academia sinica\\
 Taiwan\\
 \emph{slt.cmu@gmail.com} \and Chun-Shu Chen\\
Institute of Statistics and \\
Information Science\\
National Changhua University \\ of Education\\
Taiwan\\
\emph{cschen@cc.ncue.edu.tw} \and Yu-Fen Li\\
 Department of Public Health\\
 China Medical University\\
 Taiwan\\
\emph{yufenli@mail.cmu.edu.tw}\and Jin-Hua Chen\\
Graduate Institute of Data Science\\
Taipei Medical University\\
 Taiwan\\
\emph{jh\_chen@tmu.edu.tw}}
\maketitle
\begin{abstract}
\textbf{Objectives:} This study provides an effective model selection method based on the empirical likelihood approach for constructing summary receiver operating characteristic (sROC) curves from meta-analyses of diagnostic studies.

\textbf{Methods:} We considered models from combinations of family indices and specific pairs of transformations, which cover several widely used methods  for bivariate summary of sensitivity and specificity. Then a final model was selected using the proposed empirical likelihood method.  Simulation scenarios were conducted based on different number of studies and different population distributions for the disease and non-disease cases. The performance of our proposal and other model selection criteria
was also compared. 

\textbf{Results:} Although parametric likelihood-based methods are often applied in practice due to its asymptotic property, they fail to consistently choose appropriate models for summary under the limited number of studies. For these situations, our proposed method almost always performs better.

\textbf{Conclusion:} 
When the number of studies is as small as 10 or 5, we recommend choosing a summary model via the proposed empirical likelihood method.

\end{abstract}

\section{Introduction}
Summarizing information on test performance metrics, such as sensitivity, specificity, and diagnostic odds ratios (DOR), is an important part of a systematic review of a medical test performance on clinical outcomes. Through a meta-analysis on clinical studies of diagnostic tests, we may investigate hypotheses about the test performance that cannot be answered by an individual study. \citet{sotiriadis2016synthesizing} recommended a guideline for systematic review 
of diagnostic test accuracy studies, as a counterpart to  Cochrane Handbook (\citealp{higgins2011cochrane}) and PRISMA (\citealp{shamseer2015preferred}) widely used in general systematic review. 

Most diagnostic tests are used to separate patients into two groups as test positive and test negative (i.e., T+/T-). The fundamental diagnostic result can be a real value (continuous output), in which case the classification boundary between the two groups must be determined by a threshold value, usually based on a receiver operating characteristic (ROC) curve  (\citealp{ref01}). Accordingly, there are four possible outcomes from a dichotomized test when a gold standard is available. If the true disease status of a subject is positive (D+), a T+ classification is called a true positive (TP), while a T- result is called a false negative (FN). Conversely, given a true negative disease status of a subject (D-), a T- classification results in a true negative (TN) and a T+ result gets a false positive (FP).

In such cases that a gold standard exists, test accuracy is estimated as the proportion of diseased individuals to be ``test positive"  (sensitivity) and of non-diseased individuals to be ``test negative"  (specificity); see \citet{ref02}, \citet{ref03}, and \citet{ref04}. For systematic reviews of dichotomous diagnostic studies, we have ``data" merely consisting of numbers of nTP, nFN, nFP, and nTN for each study involved, summing up the number of subjects classified as TP, FN, FP, or TN, respectively. The corresponding results are usually reported based on a particular threshold, as used in clinical practice (\citealp{ref05}). It is improper to simply use the sums across studies of these four numbers to derive summary estimates of sensitivity, specificity and DOR, where the summary statistics would be dominantly affected by several studies in the largest study sizes.

Another naive summary is to pool sensitivity and specificity separately using standard meta-analyses for proportions. However, \citet{ref06} and \citet{ref07} showed that sensitivity and specificity are often negatively correlated, usually because of different thresholds among studies to define T+ and T-. Even though this ``separate summary"  method is sometimes recommended (e.g., \citealp{ref08}, \citealp{ref09}), ignoring the correlation between sensitivity and specificity would result in a biased inference or even misleading in the claims.

Ideally, each study has its own (empirical) ROC curve, and a summary ROC curve provides an overall description of the test performance, such as the problems dealt with in \citet{ref10}. But as is often the case, many studies reported merely a table consisting of nTP, nFN, nFP, and nTN. It will be hard to distinguish the following three sources of uncertainty involved in observed sensitivities and specificities across studies: (a) the dispersion within a study due to sampling variability, (b) heterogeneity from varying cutoff values between studies, or (c) different characteristics of populations for individual studies; refer to \citet{ref05}, \citet{ref08} and \citet{ref11} for detailed discussions.  Consequently, it is almost impossible to recover each individual ROC curve based on the limited data points without additional assumptions.

 \citet{ref08} and \citet{ref09} discussed that helpful ways about summarizing medical test studies include ``separate summary", ``summary point", and ``summary line". Some advised procedures regarding when to use which kind of summary representations were also provided in their studies. These procedures, however, partly rely on the effectiveness of the bivariate ROC model of \citet{ref12}, which allows for either a ``summary point"  or a ``summary line".  In fact, various models have been proposed in the literature for a meaningful summary ``point"  or ``line"  across all studies (e.g., \citealp{ref07}, \citealp{ref12}, \citealp{ref13}, and \citealp{ref14}).   \citet{ref12} and \citet{ref13} have become almost the \textit{de facto} standard for a ``summary point"  or a ``summary line", and \citet{harbord2006unification} showed their equivalence when  no covariates are included. If more studies are available (usually with a number larger than 30),  some sophisticated extensions attempt to incorporate other sources of heterogeneities, such as disease prevalence (\citealp{chu2009meta}), latent subgroups (\citealp{schlattmann2015mixture})  or measurement errors (\citealp{guolo2017double}).  

In view of  so many alternative methods, a natural and important question is how to select a suitable model. Recently, \citet{ref15} tried to integrate a wide range of models into a unified parametric linear mixed model framework after transformation, upon which the likelihood-maximizing approach can be utilized to estimate model parameters. It covers  \citet{ref11}, \citet{ref12}, \citealp{ref13}, and \citealp{ref14} as special cases. Furthermore, some likelihood-based criteria can be used for selecting a ``best"  model, among which the Akaike information criterion (AIC) with nice asymptotic properties is the most often used method; see \citet{ref16} for more details. When a set of candidate models is considered, we may choose the model with the smallest AIC value, and then make statistical inference based on it.

Nevertheless, the practical essence of such meta-analyses is restricted to a small sample size $N$ more often than not. Note that $N$ is the number of studies under consideration. For the same medical test design, there are usually not many compatible studies and hence the number of data points is too limited to apply the relative asymptotic theories. \citet{ref17}, \citet{ref18}, and \citet{ref19} proposed conditional AIC (cAIC)  in linear mixed models, which is a tailored model selection method for small $N$. 
In addition, an empirical likelihood (EL) method analogous to \citet{ref20} can also be used for small $N$ in practice. In this study, we focus on the issue of model selection for "summary line" situations. The key questions in this article include: (a) whether  AIC  gives an acceptable result, (b) which selection criterion (e.g., AIC, cAIC, or EL) has better performance, and (c) does there exist a criterion that performs satisfactorily under various situations especially when $N$ is small?

The rest of the article is organized as follows. Section 2 first reviews several commonly used models  and then describes some existing model selection criteria, followed by our proposed criterion. The effectiveness of our proposal 
is shown through simulations comparing to other criteria
in Section 3. An example of its application to colorectal cancer detection  
is given in Section 4. Finally, we conclude with Section 5. 

\section{Method}

\subsection{Two families of models and their special cases} \label{sect: model-review}

This section briefly describes the two model families under $t_\alpha$  transformation as illustrated in \citet{ref15} and the relations with other approaches. The details of models can be found in the original literature. 

\citet{ref15} introduced a class of monotonic transformation functions controlled by $\alpha \in [0,1]$  for  $x \in [0,1]$  given as 
\[
t_{\alpha}(x) = \alpha \log(x)-(2-\alpha) \log(1-x).
\]
Let $p_i$ and $q_i$ be the unobserved true sensitivity and false positive rate (1-specificity) for the $i$-th study, respectively. With a pair of transformation parameters $(\alpha_p, \alpha_q)^\prime$, the two transformed variables  $\left( t_{\alpha_p}(p_i), t_{\alpha_q}(q_i) \right)^\prime$ are then assumed to follow a bivariate normal distribution with mean  $\bm{\mu}=(\mu_p,\mu_q)^\prime$ and covariance matrix
\begin{equation}
\bm{\Sigma}_{i}=\left(\begin{array}{cc}
\sigma_{p}^{2} & \sigma\\
\sigma & \sigma_{q}^{2}
\end{array}\right)+\left(\begin{array}{cc}
d_{ip}^{2} & 0\\
0 & d_{iq}^{2}
\end{array}\right).
\label{eq: mixed covariance}
\end{equation}
Following \citet{ref15}, we consider two families of models by setting $d_{ip}^{2}$ and $d_{iq}^{2}$ in (\ref{eq: mixed covariance}) to different values.  The first family of models
uses fixed $d_{ip}^{2}=d_{iq}^{2}=0$  while  the second family of models takes study heteroscedasticity into account with $d_{ip}^{2}=\widehat{\mathrm{Var}}\left( \frac{\mathrm{nTP}_i}{\mathrm{nTP}_i+\mathrm{nFN}_i} \right)$ and $d_{iq}^{2}=\widehat{\mathrm{Var}}\left( \frac{\mathrm{nFP}_i}{\mathrm{nTN}_i+\mathrm{nFP}_i} \right)$, i.e., estimated variances of sensitivity and specificity for individual studies.

\citet{ref15} pointed out that $t_{\alpha}$ respectively corresponds
to logit transformation with $\alpha=1$ and log transformation with
$\alpha=2$. Moreover, $t_{\alpha}$ is also approximately proportional to the
complementary logarithmic function when $\alpha$ is around 0.6. On
the other hand, $t_{\alpha}(x)$ can be regarded as $\log(1-x)$ and
complementary $\log(1-x)$ if $\alpha=0$ and $\alpha=1.4$, respectively. 

When $\alpha_{p}=\alpha_{q}=2$, the first family of models corresponds
to the Lehmann family or proportional hazard models of \citet{ref14}.
If $\textrm{logit}(p_{i})$ and $\textrm{logit}(q_{i})$ are assumed
to follow a bivariate normal distribution, the first family of models
with $\alpha_{p}=\alpha_{q}=1$ coincides with the summary ROC method
of \citet{ref07} but based on different parameterizations. Hereafter,
the model corresponding to \citet{ref07} is called MSL method. 

When $\alpha_{p}=\alpha_{q}=1$, the second family of models is equivalent
to bivariate models of \citet{ref12} and the hierarchical summary
ROC method (HSROC) of \citet{ref13}. Furthermore, the second family
of models approaches the complementary  logarithmic models of \citet{ref11}
when $\alpha_{p}=\alpha_{q}=1.4$. Apart from $(\alpha_{p},\alpha_{q})^{\prime}$,
there are five common parameters $\bm{\theta}=(\mu_{p},\mu_{q},\sigma_{p}^{2},\sigma_{q}^{2},\sigma)^{\prime}$
involved in each family. To estimate model parameters, maximum likelihood
(ML) or restricted maximum likelihood (REML) methods can be used.
For a larger $N$, $(\alpha_{p},\alpha_{q})^{\prime}$ can also be estimated
as additional parameters. 

\subsection{Existing model selection criteria and our proposal \label{sect:criteria}}

The work of \citet{ref15} reviewed in the previous subsection generalized several widely used models. For $N\geq20$,
they also showed that it is possible to recover $(\alpha_{p},\alpha_{q})^{\prime}$
by treating it as free parameters. Nevertheless, they admitted that
it is hard to estimate $(\alpha_{p},\alpha_{q})^{\prime}$ for $N\leq10$,
and they suggested these two quantities should be fixed for a small
$N$. In practice, treating $(\alpha_{p},\alpha_{q})^{\prime}$ as
fixed or free parameters does not respectively make the problem easier
or harder; an analyst still needs a way to determine suitable values
of $(\alpha_{p},\alpha_{q})^{\prime}$ for transformation. Additionally,
although they proposed two useful families of models, little has been
known about how to select among them especially when the sample size
$N$ is small. Therefore, a good model selection strategy is important. 

Model selection can be viewed as a selection of both the model assumptions
and the estimated parameters, which amounts to a choice of underlying
probabilistic mechanism. Most of works for model selection (or variable
selection) in linear regression and generalized linear models have
been studied extensively (\citealp{ref16}, \citealp{ref21}, and
\citealp{ref22}). Unfortunately the models considered here have no variables
to be selected and the key structure, the covariance matrix $\bm{\Sigma}_{i}$,
is heavily affected by $(\alpha_{p},\alpha_{q})^{\prime}$. Selection
of the covariance structure in a linear mixed model is still a very
open research area. Yet, the two families in \citet{ref15} are linear
mixed models only after transformation, which raises another challenge
for us. Therefore, the results would be doubtful if one directly applies
the existing model selection methods to select among the two families. 

In what follows, a ``model'' indicates a triplet of  $\alpha_{p}$, $\alpha_{q}$, and an index of  family
(i.e., 1st or 2nd), so $\alpha_{p}$ and $\alpha_{q}$ are no longer
free parameters. We shall consider several model selection criteria,
and compare their performance based on simulations. The first one
is AIC (\citealp{ref23}), which was also inspected in \citet{ref15}.
Let $l_{M}(\hat{\bm{\theta}})$ be the log-likelihood for a model
$M$ and $\hat{\bm{\theta}}$ be the corresponding estimates of model
parameters. Then AIC is defined as $-2l_{M}(\hat{\bm{\theta}})+2k$
with $k$ being the number of parameters in the model $M$. Since
each model has 5 parameters, selecting the minimum AIC model amounts
to choosing the model having the largest $l_{M}(\hat{\bm{\theta}})$,
where $\hat{\bm{\theta}}$ is obtained by ML or REML. Note that AIC
has been shown to have nice asymptotic properties for model selection
(\citealp{ref16}), but the focus in this work is the small $N$ problem.
Although \citet{ref24} is a corrected version of AIC with penalty
$2k(k+1)/(N-k-1)$ instead of $2k$ for small $N$, it selects an
identical model as AIC for models considered here. 

The second and third one follow the conditional AIC  studied in \citet{ref17},
\citet{ref18}, and \citet{ref19}.
Let $\bm{y}_{i}=\left(\frac{\textrm{nTP}_{i}}{\textrm{nTP}_{i}+\textrm{nFN}_{i}},\frac{\textrm{nFP}_{i}}{\textrm{nTN}_{i}+\textrm{nFP}_{i}}\right)^{\prime}$;
$i=1,2,\ldots,N$ be the vector of observed sensitivities and 1-specificities. For
a specific model $M$, define $\hat{\bm{z}}_{i}(M)=\textrm{E}_{M}(\bm{z}_{i}|\hat{\bm{\theta}},\bm{Z})$,
where 
\[
\bm{z}_{i}=\bm{t}(\bm{y}_{i})\equiv\left(t_{\alpha_{p}}\left(\frac{\textrm{nTP}_{i}}{\textrm{nTP}_{i}+\textrm{nFN}_{i}}\right),\: t_{\alpha_{q}}\left(\frac{\textrm{nFP}_{i}}{\textrm{nTN}_{i}+\textrm{nFP}_{i}}\right)\right)^{\prime},
\]
and $\bm{Z}=(\bm{z}_{1}^{\prime},\ldots,\bm{z}_{N}^{\prime})^{\prime}$.
Thus $\hat{\bm{z}}_{i}(M)$ is the empirical best linear unbiased
predictor of $\bm{t}(\bm{\mu}_{i})\equiv\left(t_{\alpha_{p}}\left(p_{i}\right),\: t_{\alpha_{q}}\left(q_{i}\right)\right)^{\prime}$
based on the model $M$. Then the conditional AIC (cAIC)
is defined as $l_{M}^{*}(\hat{\bm{\theta}})$+ penalty, where $l_{M}^{*}(\hat{\bm{\theta}})=l_{M}(\hat{\bm{\theta}})-l_{M}(\hat{\bm{\theta}|}\hat{\bm{Z}}(M))$
with $l_{M}(\hat{\bm{\theta}|}\hat{\bm{Z}}(M))$ being the log-likelihood
for a model $M$ evaluated at $\hat{\bm{\theta}}$ and the observations
$\bm{Z}$  replaced by $\hat{\bm{Z}}(M)\equiv\left(\hat{\bm{z}}_{1}(M)^{\prime},\ldots,\hat{\bm{z}}_{N}(M)^{\prime}\right)^{\prime}$,
and the penalty in cAIC  was discussed in \citet{ref17},
\citet{ref18}, and \citet{ref19}. In particular, \citet{ref17}
 assumed $\bm{\theta}$ to be known, while \citet{ref18} and \citet{ref19}
took the uncertainty of estimation into consideration. The major difference
between \citet{ref18} and \citet{ref19} is that the former calculated
the penalty approximately, while the latter provided an exact method.
In our simulation studies, we will compare the performance of  
\citet{ref17} and \citet{ref19}, and refer to them as cAIC-VB and
cAIC-GK, respectively. Also note that a model in the first family
does not consider the random effect, hence cAIC reduces to AIC in
this case.

The fourth criterion is EL approach (\citealp{ref20}), which was primarily a method for constructing a confidence region for mean parameters, and \citet{ref25} pointed out  its connection to goodness-of-fit measures. Denote $\bm{t}^{-1}(\bm{x})\equiv\left(t_{\alpha_{p}}^{-1}\left(x_p\right),\: t_{\alpha_{q}}^{-1}\left(x_q\right)\right)^{\prime}$ for an arbitrary  $\bm{x}=(x_p,x_q)^\prime$, and hence $\bm{t}^{-1}(\bm{\mu}_{M})\equiv\left(t_{\alpha_{p}}^{-1}\left(\mu_{p}\right),\: t_{\alpha_{q}}^{-1}\left(\mu_{q}\right)\right)^{\prime}$
is the (back-transformed) mean of summarized sensitivity and 1-specificity
for a model $M$. The empirical likelihood for a model having
$\bm{\mu}_{M}=\left(\mu_{p},\mu_{q}\right)^{\prime}$ as mean parameters
is given by
\begin{equation}
L(\bm{\mu}_{M})=\max\prod_{i=1}^{N}w_{i}(\bm{\mu}_{M})  \label{eq: emplik}
\end{equation}
under the constraints $\sum_{i}w_{i}(\bm{\mu}_{M})\left(\bm{y}_{i}-\bm{t}^{-1}(\bm{\mu}_{M})\right)=\bm{0}$,
$w_{i}(\bm{\mu}_{M})>0$, and $\sum_{i}w_{i}(\bm{\mu}_{M})=1$. The
empirical likelihood for the saturated model is 
\[
L(\bm{\mu}^{*})=\max\prod_{i=1}^{N}w_{i}
\]
under the constraints $w_{i}>0$, and $\sum_{i}w_{i}=1$. To assess the hypothesis that $\bm{\mu}_{M}$ is the mean
of $N$ independent data $\bm{y}_{1},\ldots,\bm{y}_{N}$,
we should first find the weight $w_{i}(\bm{\mu}_{M})$ of each datum with (\ref{eq: emplik}).
Then, 
\[
R=-2\log\left(\frac{L(\bm{\mu}_{M})}{L(\bm{\mu}^{*})}\right)=-2\sum_{i=1}^{N}\log(w_{i}(\bm{\mu}_{M}))+2N \log(N)
\]
can be obtained, where $L(\bm{\mu}^{*})=N^{-N}$ is a constant and
$R$ has a chi-square limiting distribution with a degree of freedom equal to
the rank of $\textrm{Var}(\bm{y})$ (\citealp{ref20}). Thus, a larger value of $R$ indicates a model's deficiency. We refer to this empirical likelihood method as  EL-fix.

Note that   EL-fix cannot differentiate the covariance
structure with merely $\bm{\mu}_{M}$. We propose a simple modification as the f{}ifth criterion in the following. 
Based on an idea similar to aforementioned cAIC, we incorporate the information
of $\hat{\bm{y}}_{i}(M)\equiv\bm{t}^{-1}\left(\hat{\bm{z}}_{i}(M)\right)$;
$i=1,2,\ldots,N$, into the above empirical likelihood
method.  Specifically,  for calculating (\ref{eq: emplik}) the original constraint $\sum_{i}w_{i}(\bm{\mu}_{M})\left(\bm{y}_{i}-\bm{t}^{-1}(\bm{\mu}_{M})\right)=0$
remains the same for the first family, but we simply replace it with $\sum_{i}w_{i}(\bm{\mu}_{M})\left(\bm{y}_{i}-\hat{\bm{y}}_{i}(M)\right)=0$
for a model $M$ in the second family. Our modification is referred to
as EL-blup.  We will show the effectiveness of our proposed
method through simulation studies.

\section{Simulation Studies}

\subsection{Setup}

We shall compare the criteria described in the last section via simulation studies.
It is not fair to conduct simulations from any model for
meta-analysis in Section \ref{sect: model-review}. Generating data from a certain model
would be in favor of a specific approach, e.g., nTP and nFP come from
a bivariate binomial distribution, or logit(sensitivity) and logit(1-specificity)
come from a bivariate normal distribution. Instead, we imitate a typical data
collection process, and set up simulations similar to common demonstrations
among those methodologies for the ROC curve of a single study as in
\citet{ref26}, \citet{ref27}, or \citet{ref28}. 

We consider meta-analyses
of $N=5$ or $10$ primary studies in the diagnostic test, and for
simplicity, candidate models are restricted to those  within $(\alpha_{p},\alpha_{q})^{\prime}\in\{0,0.6,1,1.4,2\}^{2}$
in combination with the two model families. Therefore, there are $50$
candidate models under this setting. To generate data for the $i$-th study,
the primary test values of  $m_{0i}$ non-disease  participants are
drawn independently and identically from a distribution $F_{0}$, and the values of  $m_{1i}$ diseased 
participants are from $F_{1}$, where $m_{0i}$ and $m_{1i}$ are
integers sampled from Poisson distributions with means 160 and 40,
respectively. Then a threshold is determined by maximizing Youden's
index (\citealp{ref29}) for the $m_{0i}+m_{1i}$ participants, and 
we obtain the corresponding $\bm{x}_{i}=(\textrm{nTP}_{i},\textrm{nFN}_{i},\textrm{nFP}_{i},\textrm{nTN}_{i})^{\prime}$. Based on $\bm{x}_{i}$; $i=1,2,\ldots,N$, the standard 
estimation procedures for a model (a triplet of $\alpha_{p}$, $\alpha_{q}$
and index of family) is applied, and some competing model selection
criteria introduced in Section \ref{sect:criteria} will also be used. For each criterion,
a ``best\textquotedblright{} model was chosen, i.e., a model with
the smallest criterion\textquoteright{}s value is selected among
all the candidate models. 

We assess a selection criterion as follows. Let $C$ be the theoreticalROC curve in $(0,1)\times(0,1)$ space, and $A$ be the corresponding
area under the curve (i.e., AUC). For a model $M$, let $C_{M}^{*}$
and $A_{M}^{*}$ be the estimates of $C$ and $A$, respectively.
Then, $C_{M}^{*}$ and $A_{M}^{*}$ are used for assessment based on the following
four measures,
\begin{enumerate} 
\item RMSE($A_{M}^{*}$): rooted mean squared error of
$A_{M}^{*}$, 
\item rank1: the ascending ranking of RMSE($A_{M}^{*}$)
among the 50 candidate models, 
\item MIAE($C_{M}^{*}$): the mean integrated
absolute deviation between $C_{M}^{*}$ and $C$, 
\item rank2: the ascending
ranking of MIAE($C_{M}^{*}$) among the 50 candidate models. 
\end{enumerate}

The simulation experiment is replicated 500 times for each combination of $N$,
$F_{0}$, and $F_{1}$, where $F_{0}$ and $F_{1}$ are considered
under the following four scenarios: 
\begin{description}
\item  [(LD)] logistic distribution with location and scale parameters (0,1)
for $F_{0}$ and (1.8,1.2) for $F_{1}$,
\item [(ND)]  normal distribution with mean and standard deviation parameters
(0,1) for $F_{0}$ and (1.5,1.2) for $F_{1}$,
\item [(SND)] skew normal distribution with location, scale and shape parameters
(0,1,1) for $F_{0}$ and (0.25,2,5) for $F_{1}$,
\item [(TND)] truncated normal distribution  with mean and standard deviation
parameters (0,1) for $F_{0}$ and (1,1.25) for $F_{1}$, and the
truncated minimum and maximum are a standard deviation from the mean. 
\end{description}
In addition to the most popular distribution ND, LD is a heavy tail
distribution, while TND is a short tail distribution and SND is asymmetric.
These distributions are used to generate participants\textquoteright{}
situations under study and to test the performance of various model
selection criteria. 

\subsection{Results}

Since \citet{ref15} concluded that the ML estimator of the covariance
is always biased, we shall merely report results based on REML estimators.
In fact, results based on ML estimator give the same conclusion. For
reference of the upper and lower bounds, we also calculate the four
assessment measures for ``AIC-noJ\textquotedblright{} and ``BEST\textquotedblright{},
where AIC-noJ is the criterion similar to AIC but not using the Jacobian
of the transformation, and BEST collects the model having the smallest
MIAE($C_{M}^{*}$) value among the 50 candidate models for each replication.
As expected, AIC-noJ performs the worst while BEST is superior to
the others. Also note that AUC values can be close even if two $C_{M}^{*}$'s
shapes are very different, and thus comparison of $C_{M}^{*}$ is
more meaningful for ``summary line\textquotedblright{} situation. 

Tables 1 and 2 summarize the performance of different criteria for
$N=5$ and $N=10$, respectively. It is obvious that our proposed
method, EL-blup, always holds the best two places for RMSE($A_{M}^{*}$),
and outperforms others for MIAE($C_{M}^{*}$). In contrast, AIC tends
to choose worse models in many scenarios, and cAIC corrects it to
some extent. Note that a random selecting mechanism would result in
rankings of RMSE($A_{M}^{*}$) and MIAE($C_{M}^{*}$) with an average
value of 25.5. In practice, we have little knowledge about the underlying
true distributions for $F_{0}$ and $F_{1}$, so a stable and robust
method is critical. We notice that our proposed method, EL-blup, is
the only criterion steadily beating a random selection.

\begin{table}
\caption{Comparison between several model selection criteria for four different
population distributions in $N=5$ cases (the values in parentheses
are the corresponding standard errors).}
\scalebox{0.85}{
\centering{}%
\begin{tabular}{ccccccccc}
\hline 
Distributions & Measures & AIC-noJ & AIC & cAIC-VB & cAIC-GK & EL-fix & EL-blup & BEST\tabularnewline
\hline 
 & \multirow{2}{*}{RMSE} & 10.98\% & 4.18\% & 6.63\% & 6.98\% & 8.44\% & 5.57\% & 3.89\%\tabularnewline
 &  & (0.12\%) & (0.19\%) & (0.21\%) & (0.15\%) & (0.17\%) & (0.18\%) & (0.13\%)\tabularnewline
 & \multirow{2}{*}{Rank1} & 47.98  & 16.27  & 25.73  & 29.09  & 37.28  & 21.48  & 15.43 \tabularnewline
LD(0,1) vs. &  & (0.24) & (0.94) & (1.04) & (0.5) & (0.77) & (0.73) & (0.63)\tabularnewline
LD(1.8,1.2) & \multirow{2}{*}{MAIE} & 11.04\% & 8.18\% & 8.52\% & 8.13\% & 9.24\% & 7.71\% & 5.12\%\tabularnewline
 &  & (0.11\%) & (0.17\%) & (0.17\%) & (0.18\%) & (0.16\%) & (0.2\%) & (0.14\%)\tabularnewline
 & \multirow{2}{*}{Rank2} & 44.50  & 26.25  & 29.14  & 25.08  & 34.04  & 20.99  & 1.00 \tabularnewline
 &  & (0.61) & (0.81) & (0.76) & (0.64) & (0.86) & (0.82) & (0)\tabularnewline
\hline 
 & \multirow{2}{*}{RMSE} & 7.55\% & 4.07\% & 4.7\% & 4.85\% & 5.99\% & 4.13\% & 2.95\%\tabularnewline
 &  & (0.09\%) & (0.23\%) & (0.16\%) & (0.13\%) & (0.12\%) & (0.14\%) & (0.11\%)\tabularnewline
 & \multirow{2}{*}{Rank1} & 46.12  & 21.00  & 25.14  & 28.04  & 36.92  & 22.91  & 17.14 \tabularnewline
ND(0,1) vs. &  & (0.45) & (1.02) & (1.01) & (0.68) & (0.75) & (0.75) & (0.58)\tabularnewline
ND(1.5,1.2) & \multirow{2}{*}{MAIE} & 7.67\% & 7.51\% & 7.31\% & 6.65\% & 6.98\% & 6.24\% & 3.85\%\tabularnewline
 &  & (0.09\%) & (0.22\%) & (0.21\%) & (0.2\%) & (0.16\%) & (0.22\%) & (0.11\%)\tabularnewline
 & \multirow{2}{*}{Rank2} & 40.31  & 32.96  & 32.58  & 22.16  & 32.09  & 20.27  & 1.00 \tabularnewline
 &  & (0.81) & (0.8) & (0.79) & (0.77) & (0.88) & (0.85) & (0)\tabularnewline
\hline 
 & \multirow{2}{*}{RMSE} & 7.78\% & 4.46\% & 4.93\% & 4.83\% & 6.22\% & 4.33\% & 3.63\%\tabularnewline
 &  & (0.12\%) & (0.2\%) & (0.16\%) & (0.14\%) & (0.14\%) & (0.15\%) & (0.11\%)\tabularnewline
 & \multirow{2}{*}{Rank1} & 47.22  & 24.98  & 27.51  & 27.34  & 38.02  & 23.33  & 21.66 \tabularnewline
SND(0,1,1) vs.  &  & (0.46) & (0.97) & (0.95) & (0.69) & (0.72) & (0.69) & (0.56)\tabularnewline
SND(0.25,2,5) & \multirow{2}{*}{MAIE} & 8.21\% & 7.81\% & 7.64\% & 6.57\% & 6.68\% & 6.34\% & 4.47\%\tabularnewline
 &  & (0.12\%) & (0.17\%) & (0.16\%) & (0.16\%) & (0.13\%) & (0.2\%) & (0.11\%)\tabularnewline
 & \multirow{2}{*}{Rank2} & 37.01  & 35.94  & 31.69  & 18.70  & 25.69  & 16.03  & 1.00 \tabularnewline
 &  & (0.82) & (0.86) & (0.91) & (0.76) & (0.92) & (0.83) & (0)\tabularnewline
\hline 
 & \multirow{2}{*}{RMSE} & 7.21\% & 6.35\% & 5.56\% & 5.11\% & 6.03\% & 4.78\% & 3.48\%\tabularnewline
 &  & (0.14\%) & (0.29\%) & (0.23\%) & (0.17\%) & (0.15\%) & (0.19\%) & (0.13\%)\tabularnewline
 & \multirow{2}{*}{Rank1} & 40.78  & 34.61  & 28.15  & 26.45  & 29.94  & 23.58  & 19.97 \tabularnewline
TND(0,1) vs.  &  & (0.73) & (0.95) & (0.88) & (0.86) & (0.86) & (0.85) & (0.66)\tabularnewline
TND(1,1.25) & \multirow{2}{*}{MAIE} & 7.39\% & 9.39\% & 8.27\% & 7.04\%  & 6.71\%  & 6.36\%  & 4.36\%\tabularnewline
 &  & (0.16\%) & (0.24\%) & (0.23\%) & (0.26\%) & (0.2\%) & (0.17\%) & (0.16\%)\tabularnewline
 & \multirow{2}{*}{Rank2} & 28.17  & 34.86  & 29.59  & 19.92  & 18.06  & 17.24 & 1.00 \tabularnewline
 &  & (0.87) & (0.94) & (0.99) & (0.91) & (0.8) & (0.84) & (0)\tabularnewline
\hline 
\end{tabular}
}
\end{table}

\begin{table}
\caption{Comparison between several model selection criteria for four different
population distributions in $N=10$ cases (the values in parentheses
are the corresponding standard errors).}

\scalebox{0.85}{
\centering{}%
\begin{tabular}{ccccccccc}
\hline 
Distributions & Measures & AIC-noJ & AIC & cAIC-VB & cAIC-GK & EL-fix & EL-blup & BEST\tabularnewline
\hline 
 & \multirow{2}{*}{RMSE} & 11.09\% & 4.6\% & 5.88\% & 6.04\% & 8.38\% & 5.56\% & 3.00\%\tabularnewline
 &  & (0.08\%) & (0.19\%) & (0.18\%) & (0.17\%) & (0.13\%) & (0.17\%) & (0.09\%)\tabularnewline
 & \multirow{2}{*}{Rank1} & 47.63  & 16.80  & 22.10  & 23.89  & 36.68  & 19.83  & 12.02 \tabularnewline
LD(0,1) vs. &  & (0.17) & (0.85) & (0.87) & (0.82) & (0.61) & (0.7) & (0.44)\tabularnewline
LD(1.8,1.2) & \multirow{2}{*}{MAIE} & 11.1\% & 6.83\% & 7.20\% & 6.94\% & 8.63\% & 6.81\% & 4.04\%\tabularnewline
 &  & (0.08\%) & (0.14\%) & (0.12\%) & (0.11\%) & (0.13\%) & (0.16\%) & (0.09\%)\tabularnewline
 & \multirow{2}{*}{Rank2} & 46.90  & 22.50  & 24.90  & 24.06  & 34.20  & 20.58  & 1.00 \tabularnewline
 &  & (0.23) & (0.75) & (0.69) & (0.54) & (0.73) & (0.8) & (0)\tabularnewline
\hline 
 & \multirow{2}{*}{RMSE} & 7.80\% & 3.66\% & 4.44\% & 4.53\% & 5.73\% & 4.13\% & 2.36\%\tabularnewline
 &  & (0.07\%) & (0.12\%) & (0.14\%) & (0.11\%) & (0.09\%) & (0.14\%) & (0.06\%)\tabularnewline
 & \multirow{2}{*}{Rank1} & 47.50  & 21.96  & 25.23  & 26.86  & 35.82  & 23.94  & 15.76 \tabularnewline
ND(0,1) vs. &  & (0.27) & (0.88) & (0.96) & (0.6) & (0.64) & (0.76) & (0.41)\tabularnewline
ND(1.5,1.2) & \multirow{2}{*}{MAIE} & 7.79\% & 6.26\% & 6.03\% & 5.36\% & 6.11\% & 5.29\% & 3.10\%\tabularnewline
 &  & (0.07\%) & (0.14\%) & (0.12\%) & (0.13\%) & (0.13\%) & (0.18\%) & (0.06\%)\tabularnewline
 & \multirow{2}{*}{Rank2} & 45.38  & 30.92  & 29.40  & 18.96  & 30.38  & 18.76  & 1.00 \tabularnewline
 &  & (0.42) & (0.78) & (0.72) & (0.76) & (0.75) & (0.88) & (0)\tabularnewline
\hline 
 & \multirow{2}{*}{RMSE} & 7.85\% & 4.33\% & 4.66\% & 4.57\% & 6.04\% & 4.07\% & 3.21\%\tabularnewline
 &  & (0.08\%) & (0.14\%) & (0.13\%) & (0.11\%) & (0.1\%) & (0.13\%) & (0.07\%)\tabularnewline
 & \multirow{2}{*}{Rank1} & 45.78  & 24.40  & 27.20  & 26.89  & 37.38  & 22.91  & 21.01 \tabularnewline
SND(0,1,1) vs.  &  & (0.19) & (0.86) & (0.78) & (0.56) & (0.52) & (0.74) & (0.41)\tabularnewline
SND(0.25,2,5) & \multirow{2}{*}{MAIE} & 7.83\% & 6.75\% & 6.09\% & 5.28\% & 6.14\% & 5.24\% & 3.98\%\tabularnewline
 &  & (0.08\%) & (0.13\%) & (0.09\%) & (0.09\%) & (0.10\%) & (0.11\%) & (0.06\%)\tabularnewline
 & \multirow{2}{*}{Rank2} & 41.86  & 27.62  & 23.55  & 14.36  & 25.71  & 13.14  & 1.00 \tabularnewline
 &  & (0.5) & (0.98) & (0.81) & (0.78) & (0.85) & (0.7) & (0)\tabularnewline
\hline 
 & \multirow{2}{*}{RMSE} & 7.1\% & 4.72\% & 4.97\% & 4.66\% & 5.73\% & 4.57\% & 2.9\%\tabularnewline
 &  & (0.1\%) & (0.2\%) & (0.13\%) & (0.14\%) & (0.12\%) & (0.15\%) & (0.06\%)\tabularnewline
 & \multirow{2}{*}{Rank1} & 45.04  & 27.36  & 27.51  & 26.87  & 36.28  & 22.43  & 19.59 \tabularnewline
TND(0,1) vs.  &  & (0.39) & (0.9) & (0.78) & (0.86) & (0.71) & (0.94) & (0.52)\tabularnewline
TND(1,1.25) & \multirow{2}{*}{MAIE} & 7.46\% & 7.13\% & 6.35\% & 6.10\% & 5.88\% & 5.84\% & 3.84\%\tabularnewline
 &  & (0.09\%) & (0.19\%) & (0.20\%) & (0.13\%) & (0.12\%) & (0.11\%) & (0.07\%)\tabularnewline
 & \multirow{2}{*}{Rank2} & 33.12  & 27.74  & 21.61  & 20.75  & 19.66  & 18.44  & 1.00 \tabularnewline
 &  & (0.63) & (1.01) & (0.88) & (0.84) & (0.83) & (0.78) & (0)\tabularnewline
\hline 
\end{tabular}
}
\end{table}

\section{Example in real applications \label{sec:Real-Data-Application}}

In this section, we demonstrate various methods for a dataset provided in \citet{ref30}, which evaluated the ability of microRNAs in the detection of colorectal cancer. There were only 13 studies and the authors applied the MSL method and pooled these studies together in their meta-analysis. Here, we conjecture that the precision of the diagnostic tools may keep improving with time. Thus, 13 studies are separated into two groups by their published years in our analysis (i.e., year$\leq 2011$ and year $\geq 2012$). Results are shown in Figure 1.

Even though the shapes of sROC curves are very different, the values of AUC are similar. In the first group, HSROC and EL-blup gave almost identical results, while the model selected by AIC, MSL and the Lehmann model gave slightly broader confidence and prediction regions. In the second group, we notice that the model selected by AIC produced a curve close to that of the HSROC model. However, both of them failed to capture the characteristics for data points, because they gave too narrow confidence and prediction regions for sensitivity and 1-specificitiy and both models estimated very high negative correlations between them. In contrast, EL-blup, MSL and Lehmann methods gave more reasonable confidence and prediction regions. Results from the EL-blup method somewhat support our conjecture, though the precision difference between the two groups is not significant.

\begin{figure}
	\begin{centering}
         \vspace{-0.5cm}
		\includegraphics[scale=0.108]{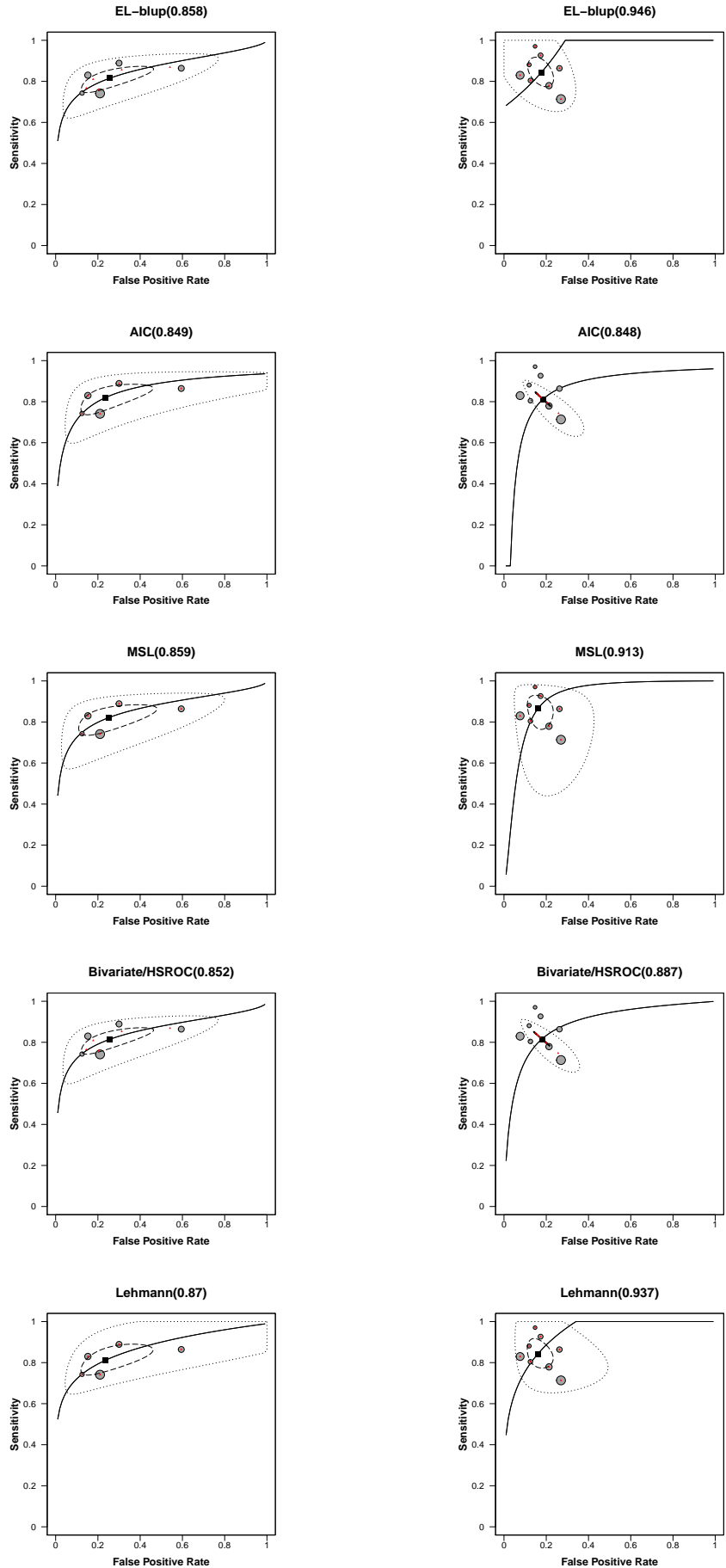}
		\par\end{centering}
	\caption{\label{fig:real-example} Summary ROC curves based on different methods for the two time periods. Left panel: year$\leq 2011$; right panel: year $\geq 2012$. Parentheses are the corresponding AUC values.}
\end{figure}

\section{Conclusion}

The model selection problem for  meta-analyses of diagnostic studies can be very difficult, not only because of the small sample size, but also due to the probabilistic mechanism of models not perfectly coinciding with the data collection process. We can almost conclude that there is no true model. 
The common criteria based on asymptotic theories do not have acceptable performance in such challenging cases. Our method can provide a more credible inference as demonstrated in the simulation studies and the real data example, even though we do not know the underlying distributions.

\bibliography{sroc}
\end{document}